\documentclass[a4,12pt]{article}
\usepackage{latexsym}
\usepackage{comment}

\usepackage[dvipdfmx]{graphicx}
\usepackage{amsmath,amssymb}
\let\du=\du                     

\newcommand{\be}{\begin{equation}}
\newcommand{\ee}{\end{equation}}
\newcommand{\nbe}{\begin{equation*}}
\newcommand{\nee}{\end{equation*}}

\newcommand{\lb}{\label}

\def\Tilde#1{\widetilde{#1}}                    

\begin{document}

\thispagestyle{empty}

{\hbox to\hsize{
\vbox{\noindent February 2017 \hfill IPMU17-0006 }}}
{\hbox to\hsize{
\vbox{\noindent  revised version \hfill }}}

\noindent
\vskip2.0cm
\begin{center}

{\large\bf Inflation from $(R+\gamma R^n-2\Lambda)$ Gravity in Higher Dimensions}
\vglue.3in

Sergei V. Ketov~${}^{a,b,c}$ and Hiroshi Nakada~${}^{a}$  
\vglue.1in

${}^a$~Department of Physics, Tokyo Metropolitan University, \\
Minami-ohsawa 1-1, Hachioji-shi, Tokyo 192-0397, Japan \\
${}^b$~Kavli Institute for the Physics and Mathematics of the Universe (IPMU),
\\The University of Tokyo, Chiba 277-8568, Japan \\
${}^c$~Institute of Physics and Technology, Tomsk Polytechnic University,\\
30 Lenin Avenue, Tomsk 634050, Russian Federation \\
\vglue.1in
ketov@tmu.ac.jp, nakada-hiroshi1@ed.tmu.ac.jp
\end{center}

\vglue.3in

\begin{center}
{\Large\bf Abstract} 
\end{center}

\noindent We propose a derivation of the inflaton scalar potential from the higher $(D)$ dimensional 
$(R+\gamma R^n-2\Lambda)$ gravity, with the new coupling constant $\gamma$ and the cosmological constant 
$\Lambda$. We assume that a compactification of extra dimensions happened before inflation, so that the inflaton scalar potential in four spacetime dimensions appears to be dependent upon the parameters $(\gamma,\Lambda ,D,n)$. We find that consistency requires $n=D/2$, while the dimension $D$ has to be a multiple of four. We calculate the potential for any $D$, and determine the values of $\gamma$ and $\Lambda$ from observations. The cases of   $D=8$ and $D=12$ are considered in more detail. Our approach results in the viable models of chaotic large-field inflation, and leads to the sharp predictions of an observable value of the tensor-to-scalar ratio $r$ of the Cosmic Microwave Background radiation.

\newpage

\section{Introduction}

The $f(R)$-gravity (with $R$ representing the spacetime curvature scalar) is the standard theoretical tool in modern cosmology, being one of the simplest approaches to modified gravity capable of
describing both cosmological inflation in the early universe and dark energy in the present universe --- see e.g., the reviews \cite{svrev,tsu,clrev,myrev,starp} and references therein.

The classical equivalence (duality) between the $f(R)$ gravity models and the scalar-tensor gravity models \cite{fmbook} is known for the long time and
has many physical applications --- see e.g., Refs.~\cite{wag,bick,whitt,jk,bc,maeda,kkw1,kw1,kw2}. In particular, the existence of the (Legendre-Weyl) equivalence transformation relating those apparently different gravity models is guaranteed by the physical conditions on the $f(R)$-function, namely, positivity of its first and second derivatives (in the proper notation). 

The simplest $f(R)$ gravity is given by the famous Starobinsky model of $(R+\gamma R^2)$ gravity with the action \cite{star1} 
 \be \label{star}
S_{\rm Starobinsky} = \frac{1}{2} \int \mathrm{d}^4x\sqrt{-g} \left[ R +\frac{1}{6M^2}R^2\right]~.
\ee
We employ the natural units $\hbar=c=1$ with the reduced Planck mass $M_{\rm Pl}=1$, and spacetime signature $(-,+,+,+)$ in four dimensions,
$D=4$. The Starobinsky model is an excellent model of inflation, in very good agreement with the Planck data \cite{Ade:2015lrj}. Actually, any viable
 inflationary model of $f(R)$ gravity must be close to the Starobinsky model (\ref{star}) in the sense of having $f(R)=R^2 A(R)$ with a slowly-varying function  $A(R)$. In particular, any $(R+\gamma R^n)$ gravity model in $D=4$ with an integer power $n$ higher than two is not viable for inflation \cite{kkw2}. 
 
The Starobinsky model is geometrical (i.e. it includes "gravity" only),
while its only real parameter $M$ can be identified with the scalaron (inflaton) mass, whose value is fixed by the observational Cosmic 
Microwave Background (CMB) data as $M=(3.0 \times10^{-6})(\frac{50}{N_e})$ where $N_e$ is the e-foldings number.  The corresponding scalar potential of the (canonically normalized) scalaron (inflaton) field $\phi$ in the dual (scalar-tensor gravity) picture is given by
\cite{jdb}
\begin{equation} \label{starpot}
V(\phi) = \frac{3}{4} M^2\left( 1- e^{-\sqrt{\frac{2}{3}}\phi }\right)^2~.
\end{equation}
This scalar potential is bounded from below (actually, is non-negative and stable), with its minimum corresponding to a Minkowski vacuum. The scalar potential (\ref{starpot}) has {\it a plateau} of a positive height, when $\phi\gg 1$, describing slow roll inflation. During this slow roll of inflaton along the plateau  the scalar potential (\ref{starpot}) can be simplified to
\begin{equation} \label{starpot2}
V(\phi) \approx V_0\left( 1- 2e^{-\alpha_s\phi }\right)~,
\end{equation}
where we have kept only the leading (exponentially small) correction to the emergent cosmological constant $V_0=\frac{3}{4} M^2$ and have introduced the notation $ \alpha_s=\sqrt{\frac{2}{3}}$. Hence, as regards the slow-roll inflation, the scalar potential  (\ref{starpot}) is  the particular case of a special class of inflationary plateau-like scalar potentials having the form
\begin{equation} \label{pclass}
V(\phi) = V_0 - V_1e^{-\alpha\phi}
\end{equation}
with {\it generic} positive real parameters $V_0$, $V_1$ and $\alpha$. The $V_1$ is obviously unimportant because it can be easily changed  to any desired value by a shift of the field $\phi$ in Eq.~(\ref{pclass}). The $V_0$ determines the scale of inflation. And the value of $\alpha$ determines the key observational parameter $r$ related to primordial gravity waves and known as the tensor-to-scalar ratio, as
\begin{equation} \lb{rbound}
r =\frac{8}{\alpha^2N_e^2}~~.
\end{equation}
The Planck data \cite{Ade:2015lrj} sets the upper bound on $r$ (with 95\% of CL) as $r<0.08$ that implies 
\begin{equation}\lb{abound}
\alpha >\frac{10}{N_e}= 0.2 \left(\frac{50}{N_e}\right)~~.
\end{equation}

As regards the other CMB spectral tilts (inflationary observables), the scalar spectral index $n_s$ and its running $dn_s/d{\ln}k$,
their values derived from the potential  (\ref{pclass}) are the same as those in the Starobinsky case, namely
\begin{equation} \label{otilts}
n_s\approx 1 - \frac{2}{N_e} \qquad {\rm and} \qquad \frac{dn_s}{d \ln k} \approx -\frac{(1-n_s)^2}{2}\approx -\frac{2}{N_e^2}~~.
\end{equation}

The scalar potentials (\ref{pclass}) with generic $\alpha$ also arise in the reconstruction of the inflaton scalar potential from the observable power spectrum of scalar perturbations, when betting on very small values of $r$ and demanding a plateau-like scalar potential.~\footnote{A.A. Starobinsky, private communication.} However, only in the case
of $\alpha=\alpha_s$ the inverse duality transformation applied to the scalar potential  (\ref{starpot}) reproduces the Starobinsky 
$(R+\gamma R^2)$ model (\ref{star}). 

In this paper we study the $(R+\gamma R^n-2\Lambda)$ gravity models in {\it higher} $(D>4)$ spacetime dimensions with the cosmological constant $\Lambda$, in an effort to derive some scalar potentials of the type (\ref{starpot}) or (\ref{pclass}), leading to the very {\it specific} values of $\alpha$  that have their origin in a higher-dimensional modified gravity. We stress that we do {\it not\/} mean a cosmological inflation in higher dimensions. We assume that our universe was born multi-dimensional, and then four spacetime dimensions became infinite, while the others curled up by some unknown mechanism {\it before inflation}. We exploit the fact that the Weyl transform, as part of the duality transformation between Jordan and Einstein frames, depends upon $D$ \cite{bc,maeda}.
 We apply the duality transformation to $f(R)$ gravity in $D$ dimensions, get the scalaron (inflaton) scalar potential, and after that dimensionally reduce it (by integrating over compact dimensions) to four (infinite) spacetime dimensions. The cosmological inflation is assumed to be taking place after compactification.

Our paper is organized as follows. In Sec.~2 we apply the Legendre-Weyl duality transformation to the $(R+\gamma R^n-2\Lambda)$ gravity model in  arbitrary ($D$) spacetime dimensions. Demanding completeness in the field space determines the allowed values of $n$. Then we perform dimensional reduction of the derived scalar-tensor gravity to four spacetime dimensions.  The scalar potential is studied in Sec.~3. In Sec.~4 we apply our results to the higher dimensions $D=8$ and $D=12$, as the simplest non-trivial examples. Sec.~5 is our conclusion.
\vglue.3in

\section{Our setup}

We denote spacetime vector indices in $D$ dimensions by capital latin letters $A,B,\ldots=0,1,\ldots, D-1$, and spacetime vector indices in four dimensions by lower case greek letters $\alpha,\beta,\ldots=0,1,2,3$. In this Section we proceed along the lines of Ref.~\cite{nk1}, though in $D$ dimensions and with arbitrary $n$.

Our starting point is the following gravitational action in a $D$-dimensional curved spacetime:
\begin{equation} \lb{action1}
 S_{\rm grav.}=\frac{1}{2\kappa^2}\int d^Dx\sqrt{-g_D}(R+\gamma R^n-2\Lambda)~~,
\end{equation}
where $\kappa>0$ is the gravitational coupling constant of (mass) dimension $\frac{1}{2}(-D+2)$,
$\gamma>0$ is the new (modified gravity) coupling constant of (mass) dimension $(-2n+2$), 
and $\Lambda$ is the cosmological constant of (mass) dimension $2$, in $D$ dimensions. A spontaneous compactification of the 
multi-dimensional gravity models (\ref{action1}) to AdS spaces and its stabilization were studied in Refs.~\cite{zhuk1,zhuk2,zhuk3} by
employing the duality transform to the scalar-tensor gravity picture in $D$ dimensions. In our paper we take a different route, by demanding the scalar potential to have a plateau of positive height, instead of its vanishing (at large field values) assumed in 
Refs.~\cite{zhuk1,zhuk2,zhuk3}.

After a substitution 
\begin{equation} \lb{subst}
 R+\gamma R^n \longrightarrow (1+B)R-\left(\frac{1}{\gamma n}\right)^{\frac{1}{n-1}}\left(\frac{n-1}{n}\right)B^{\frac{n}{n-1}}~~,
\end{equation}
where we have introduced the new scalar field $B$,  the action (\ref{action1}) takes the form
\begin{equation}  \lb{action2}
S =\frac{1}{2\kappa^2}\int d^Dx\sqrt{-g_D}\left[(1+
 B)R-\left(\frac{1}{\gamma n}\right)^{\frac{1}{n-1}}\left(\frac{n-1}{n}\right)B^{\frac{n}{n-1}}-2\Lambda\right]~~.
\end{equation}
The field $B$ enters the action (\ref{action2})  algebraically, while its "equation of motion" reads $B=\gamma n R^{n-1}$. 
After substituting the latter back into the action (\ref{action2})  we get the original action (\ref{action1}). Hence, the actions
(\ref{action1}) and (\ref{action2}) are classically equivalent. 

Next, we apply a Weyl transformation with the space-time-dependent parameter $\Omega(x)$,
\begin{equation} \lb{weylt}
 g_{AB}=\Omega^{-2}\tilde{g}_{ AB},\ \ \ \sqrt{-g}=\Omega^{-D}\sqrt{-\tilde{g}}~~,
\end{equation}
where we have introduced the new spacetime metric $\tilde{g}_{ AB}$ in $D$ dimensions. As a result of this transformation, the
corresponding scalar curvatures are related by
\begin{equation} \lb{weylr}
 R=\Omega^2[\tilde{R}+2(D-1)\Tilde{\square}f-(D-1)(D-2)\tilde{g}^{AB}f_Af_B]~~,
\end{equation}
where we have introduced the notation
\begin{equation}
 f=\ln \Omega~,\qquad f_A=\frac{\partial_A\Omega}{\Omega}~,
\end{equation}
and the covariant wave operator $\Tilde{\square}=\Tilde{D}^A\Tilde{D}_A$ in $D$ dimensions.

The Weyl-transformed (and equivalent via the field-redefinition (\ref{weylt})) action $S$ is given by
\begin{align} \lb{action3}
 S=&\frac{1}{2\kappa^2}\int d^Dx\sqrt{-\tilde{g}_{ D}}\Omega^{-D}[(1+B)\Omega^2(\tilde{R}+2(D-1)\Tilde{\square}f\nonumber\\
-&(D-1)(D-2)\tilde{g}^{AB}f_Af_B)-\left(\frac{1}{\gamma n}\right)^{\frac{1}{n-1}}\frac{n-1}{n}B^{\frac{n}{n-1}}-2\Lambda]~.
\end{align}
Hence, in order to get the corresponding action in Einstein frame, we should choose the local parameter $\Omega$ as
\begin{equation} \lb{omegaf}
\Omega^{D-2}=e^{(D-2)f}=1+B~~.
\end{equation}
We thus find 
\begin{equation}
 f=\frac{1}{D-2}\ln(1+B)
 \end{equation}
and
\begin{align} \lb{action4}
 S=&\frac{1}{2\kappa^2} \int d^Dx \sqrt{-\tilde{g}_{D}} \left[ \tilde{R}-(D-1)(D-2)\tilde{g}^{AB}\partial_Af\partial_Bf \right. \nonumber\\
 - & \left. e^{-Df} \left( \frac{1}{\gamma n}\right)^{\frac{1}{n-1}} \frac{n-1}{n}B^{\frac{n}{n-1}}- 2e^{-Df}\Lambda \right]~~.
\end{align}

As is clear from Eq.~(\ref{action4}), we should also rescale the scalar field $f$, in order to get the canonically normalized scalar kinetic terms, as
\begin{equation}\lb{canf}
\phi=\sqrt{\frac{(D-1)(D-2)}{\kappa^2}}\; f~~.
\end{equation}
As a result, in terms of the canonical scalar $\phi$, we find
\begin{equation}
 B=e^{(D-2)\kappa\phi/ \sqrt{(D-1)(D-2)}}-1~~,
\end{equation}
the scalar potential 
\begin{align} \lb{gsp}
2\kappa^2 V(\phi) =  &
 \left( \frac{1}{\gamma n}\right)^{\frac{1}{n-1}} \left( \frac{n-1}{n}\right)
 \left[ e^{(D-2)\kappa\phi/ \sqrt{(D-1)(D-2)}}-1\right]^{\frac{n}{n-1}} \times \nonumber \\
& \times  e^{-D\kappa\phi/ \sqrt{(D-1)(D-2)}}   +  2\Lambda e^{-D\kappa\phi/ \sqrt{(D-1)(D-2)}}~~,
\end{align}
and the standard scalar-tensor gravity action in Einstein frame in $D$ dimensions,
\begin{equation} \lb{action5}
 S=\frac{1}{2\kappa^2}\int
  d^Dx\sqrt{-\tilde{g}_{D}}\tilde{R}+\int d^Dx\sqrt{-\tilde{g}_{D}}\left[ -\frac{1}{2}\tilde{g}^{AB}\partial_A\phi\partial_B\phi-V(\phi)\right]~.
\end{equation}

We assume that the $D$-dimensional action (\ref{action5}) is  then "compactified" to four infinite spacetime dimensions before inflation. Applying dimensional reduction (i.e. taking all fields to be independent upon compact flat $(D-4)$ dimensions), we have
\begin{equation} \lb{dimr}
\int  d^Dx = V_{D-4} \int d^4x~,\quad \phi=\phi_4/ \sqrt{V_{D-4}}~, \quad \kappa =  \kappa_4\sqrt{V_{D-4}}~, \quad
V=V_4/V_{D-4}~,
\end{equation}
so that $\kappa\phi=\kappa_4\phi_4$ and $\kappa^2V=\kappa_4^2V_4$, where we have introduced the volume $V_{D-4}$ of compact dimensions, with the subscripts
"$4$" referring to four spacetime dimensions.~\footnote{In our notation, $\kappa_4=1/M_{\rm Pl}=1$.}

It gives rise to the standard four-dimensional action (in Einstein frame, and with a canonical scalar $\phi_4$)
\begin{equation} \lb{faction}
 S_{\rm inf.}[\tilde{g}_{4},\phi_4]=\frac{1}{2}\int d^4x\sqrt{-\tilde{g}_4}\tilde{R}_4+\int d^4x\sqrt{-\tilde{g}_{4}}\left[ -\frac{1}{2}\tilde{g}^{\mu\nu}_4
 \partial_{\mu}\phi_4\partial_{\nu}\phi_4-V_4(\phi_4)\right]
\end{equation}
that we are going to consider as our inflationary model in four spacetime dimensions. In what follows we stay in four spacetime dimensions. However, the higher dimension $D$ and the power $n$
enter the four-dimensional scalar potential $V_4(\phi_4)$ as the parameters, according to Eqs.~(\ref{gsp}) and
(\ref{dimr}) !

\section{The scalar potential}

To study our scalar potential in four spacetime dimensions, we rescale the relevant quantities by introducing the notation 
\begin{equation} \lb{n1}
\lambda=\left( \frac{n}{n-1}\right) \left(\frac{1}{\gamma n}\right)^{-\frac{1}{n-1}}2\Lambda~,
\end{equation}
\begin{equation} \lb{n2}
\tilde{\phi}=\frac{\phi_4}{\sqrt{(D-1)(D-2)}}~~, \\ 
\end{equation}
and
\begin{equation} \lb{n3}
\tilde{V}(\tilde{\phi})=\frac{2V_4(\phi_4)}{\left(\frac{1}{\gamma n}\right)^{\frac{1}{n-1}}\left(\frac{n-1}{n}\right)}~~.
\end{equation}

Then the scalar potential in $D$ dimensions takes the simple form
\begin{equation} \lb{sipot}
\tilde{V}(\tilde{\phi})=\left[ e^{(D-2)\tilde{\phi}}-1\right]^{\frac{n}{n-1}}e^{-D\tilde{\phi}}+\lambda e^{-D\tilde{\phi}}~~,
\end{equation}
where we have also used Eqs.~(\ref{gsp}) and (\ref{dimr}). Demanding this scalar potential to have {\it a plateau} of a positive height for  $\tilde{\phi}\rightarrow \infty$,
like that in Eq.~(\ref{starpot}), we get the condition  
\begin{equation}
\left[e^{(D-2)\tilde{\phi}}\right]^{\frac{n}{n-1}}e^{-D\tilde{\phi}}=1
\end{equation}
that implies ({\it cf.} Refs.~\cite{bc,zhuk3})
\begin{equation}\lb{dn}
n=\frac{D}{2}~~.
\end{equation}
Substituting it back to Eq.~(\ref{sipot}) yields the potential
\begin{equation} \lb{sipot2}
\tilde{V}(\tilde{\phi})=\left[1-e^{-(D-2)\tilde{\phi}}\right]^{\frac{D}{D-2}}+\lambda e^{-D\tilde{\phi}}~~.
\end{equation}

Let us write down the power $D/(D-2)=p/q$ in terms of mutually prime positive integers $p$ and $q$. Should $q$ be even, it leads to the obstruction $\tilde{\phi}\geq0$ of the real scalar field $\tilde{\phi}$ because its scalar potential becomes imaginary for  $\tilde{\phi}<0$.  For example, it happens when $D=6$ and $D=10$. Avoiding such situation puts a severe restriction on the allowed values of $D$ in our approach. Similarly, since $n$ is also the power of $R$ in Eq.~(\ref{action1}), and $R$ can take negative values, we conclude that $n$ must be integer and, hence, $D$ must be {\it even}.  The allowed dimensions are thus must be multiples of four, with the lowest values beyond four being $D=8$ and $D=12$.\footnote{The existence of  the "critical" dimension $D=8$ was also noticed in Ref.~\cite{zhuk3}.} These two cases are studied in more detail in the next Sec.~4.

Requiring the scalar potential to be bounded from below is needed for stability. It the limit $\tilde{\phi}\rightarrow -\infty$, the leading term in the scalar potential (\ref{sipot2}) is given by 
\begin{equation}\lb{nlim}
\lim_{\tilde{\phi}\rightarrow -\infty} \tilde{V}(\tilde{\phi})\approx
\left[(-1)^{\frac{D}{D-2}}+\lambda\right]e^{-D\tilde{\phi}}~~,
\end{equation}
so that we have to restrict the parameter $\lambda$ as
\begin{equation} \lb{lambdar}
\lambda \geq -(-1)^{\frac{D}{D-2}}~~.
\end{equation}

If we require the existence of a minimum of the scalar potential, describing the classical vacuum after inflation, we need the existence of a real (finite) solution to $\frac{d\tilde{V}}{d\tilde{\phi}}=0$. We find it at
\begin{equation}  \lb{fmin}
\tilde{\phi}_0=\frac{1}{D-2}\ln\left(1+\lambda^{\frac{D-2}{2}}\right)
\end{equation}
with
\begin{equation} \lb{pmin}
\tilde{V}(\tilde{\phi}_0)=\lambda\left(1+\lambda^{\frac{D-2}{2}}\right)^{\frac{-2}{D-2}}~~.
\end{equation}
It gives rise to a bit stronger condition, 
\begin{equation}
\lambda > (-1)^{\frac{2}{D-2}}~,
\end{equation}
and amounts to $\lambda > -1$ in the allowed dimensions.

A stronger condition arises by demanding the scalar potential minimum to correspond either a Minkowski or a de Sitter vacuum.
According to Eq.~(\ref{pmin}),  $\tilde{V}(\tilde{\phi}_0)\geq 0$ implies
\begin{equation}\lb{lambdalim}
\lambda \geq 0~.
\end{equation}
Finally, demanding the second derivative of the scalar potential at its minimum to be finite and positive or, equivalently, requiring a finite positive scalaron mass, restricts $\lambda$ by
\begin{equation}\lb{lambdafin}
\lambda > 0~,
\end{equation}
and implies {\it positive} cosmological constants in both $D$ and four dimensions, because of Eqs.~(\ref{n1}) and (\ref{pmin}).~\footnote{Unlike Refs.~\cite{zhuk1,zhuk2,zhuk3} that result in negative cosmological constants, we do not impose any conditions on compactification. The compactification should be addressed in a more fundamental framework (like supergravity or superstrings) with more fields and more couplings involved. }

Under the conditions above, with $n=\frac{D}{2}$, we find the four-dimensional scalar potential as
\begin{equation}\lb{4dspot}
V_4(\phi_4)=\left(\frac{2}{\gamma D}\right)^{\frac{2}{D-2}}\left(\frac{D-2}{2D}\right)\left(1-e^{-\sqrt{\frac{D-2}{D-1}}\phi_4} \right)^{\frac{D}{D-2}}
+\Lambda e^{-\sqrt{\frac{D^2}{(D-1)(D-2)}}\phi_4}~.
\end{equation}

Taylor expansion of the potential around its minimum at $\phi^{(0)}_4$,
\begin{equation} \lb{taylor}
V_4(\phi_4)=V_4(\phi^{(0)}_4)+\frac{1}{2}\frac{dV_4^2(\phi^{(0)}_4)}{d\phi_4^2}(\phi_4-\phi^{(0)}_4)^2+\ldots~,
\end{equation}
yields the cosmological constant $\delta$ in four dimensions as
\begin{equation} \lb{ccvalue}
V_4(\phi^{(0)}_4)=\Lambda\left[ 1+\frac{\gamma D}{2}\left(\frac{2D}{D-2}\Lambda\right)^{\frac{D-2}{2}}\right]^{\frac{-2}{D-2}}\equiv \delta,
\end{equation}
and the inflaton mass $M$ as
\begin{align}\lb{infmass}
\frac{dV^2_4(\phi^{(0)}_4)}{d\phi^2_4}=&\frac{2D}{(D-1)(D-2)}\Lambda\left[1+\frac{\gamma D}{2}\left(\frac{2D}{D-2}\Lambda\right)^{\frac{D-2}{2}}\right]^{-\frac{2}{D-2}}\times\\
\ &\times\left[\frac{\gamma D}{2}\left(\frac{2D}{D-2}\Lambda\right)^{\frac{D-2}{2}}\right]^{-1}\equiv M^2~~.\nonumber
\end{align}

Equations (\ref{ccvalue}) and (\ref{infmass}) can be considered as a system of two equations on the two parameters  $\Lambda$ and 
$\gamma$ of  our model, because the observational values of $\delta$ and $M$ are known as $\delta={\cal O}(10^{-120})$ and
$M\approx 3\times 10^{-6}$, respectively. We find
\begin{equation}\lb{ccinv}
\Lambda^{\frac{D-2}{2}}=\frac{2D}{(D-1)(D-2)}\delta^{\frac{D}{2}}M^{-2}+\delta^{\frac{D-2}{2}}
\end{equation}
and
\begin{equation}\lb{ainv}
\gamma=\frac{4}{(D-1)(D-2)}\left(\frac{2D}{D-2}\right)^{-\frac{D-2}{2}}\delta M^{-2} \left[ \frac{2D}{(D-1)(D-2)}\delta^{\frac{D}{2}}M^{-2}+\delta^{\frac{D-2}{2}}\right]^{-1}~~.
\end{equation}
Because of the tiny value of the cosmological constant $\delta$,  the solutions can be greatly simplified to 
\begin{equation} \lb{ccinv2}
\Lambda=\delta~~,
\end{equation}
as expected, and
\begin{equation}\lb{ainv2}
\gamma=\frac{4}{(D-1)(D-2)}\left(\frac{D-2}{2D}\right)^{\frac{D-2}{2}} M^{-2} \delta^{\frac{4-D}{2}}~~.
\end{equation}
As a check, when $D=4$ and $\Lambda=0$, we recover the Starobinsky model (\ref{star}) of Sec.~1, with $\gamma_4=1/(6M^2)\approx \frac{1}{54}10^{12}$.  Otherwise, for any $D$, we find
\begin{equation}\lb{alphaf}
\gamma\approx\frac{4}{9(D-1)(D-2)}\left(\frac{D-2}{2D}\right)^{\frac{D-2}{2}}10^{12+60(D-4)}~~.
\end{equation}

The parameter $\gamma$ in $D>4$ generically diverges when $\Lambda\to +0$ (unless $M^{4} \delta^{D-4}=const.>0$), which also implies $\delta\to +0$ and $M\to +\infty$ because the 2nd derivative (\ref{infmass}) of the scalar potential (\ref{4dspot}) becomes infinite at the minimum in this limit.  It is remarkable that a (finite) positive cosmological constant ensures $M$ to be finite too.

\section{Examples: $D=8$ and $D=12$}

In this Section, we specify our results to the two particular cases, having the special dimensions $D=8$ and $D=12$, respectively,
and with $\lambda\geq 0$.

\begin{itemize}
\item 
As regards $D=8$ and $n=4$, the scalar potential (\ref{sipot2}) reads
\begin{equation} \lb{spot8}
\tilde{V}(\tilde{\phi})=\left(1-e^{-6\tilde{\phi}}\right)^{\frac{4}{3}}+\lambda e^{-8\tilde{\phi}}~~.
\end{equation}
It has the absolute minimum at 
\begin{equation} \lb{min8}
\tilde{\phi}_0=\frac{1}{6}\ln\left(1+\lambda^3\right)~~,
\end{equation}
where it has a value (the cosmological constant) 
\begin{equation} \lb{vm8}
\tilde{V}(\lambda)=\lambda(1+\lambda^3)^{-\frac{1}{3}}~~.
\end{equation}

A profile of the four-dimensional inflaton scalar potential, originating from $D=8$, is given in Fig.~1.

\begin{figure}[htbp] 
\begin{center}
\includegraphics[clip, width=8cm]{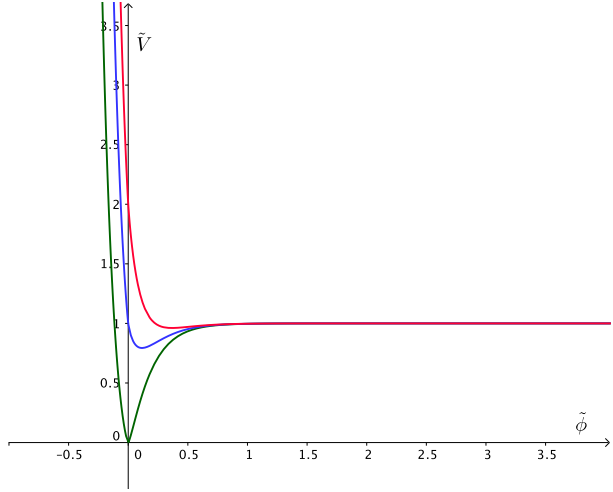}
\caption{The profile of the four-dimensional scalar potential obtained from $D=8$ dimensions. The green line 
describes the case of $\lambda=0$, the blue line is of $\lambda=1$, and the red line is of $\lambda=2$, respectively.}
\label{fig:one}
\end{center}
\end{figure}

According to Eq.~(\ref{alphaf}), the parameter $\gamma$ in $D=8$ has the value  $\gamma_8\approx\frac{1}{2^8\cdot 7}10^{252}$. Since its
(mass) dimension is $2-D=-6$, the relevant scale in $D=8$ is given by  
\begin{equation}\lb{scale8}
\gamma_8^{-1/6}\approx 3.485\cdot 10^{-42}M_{\rm Pl}~~.
\end{equation} 

\item
 As regards $D=12$ and $n=6$, the scalar potential (\ref{sipot2}) reads
\begin{equation}\lb{spot12}
\tilde{V}(\tilde{\phi})=\left(1-e^{-10\tilde{\phi}}\right)^{\frac{6}{5}}+\lambda e^{-12\tilde{\phi}}~~.
\end{equation}
It has the absolute minimum at 
\begin{equation}\lb{min12}
\tilde{\phi}_0=\frac{1}{10}\ln\left(1+\lambda^5\right)~~,
\end{equation}
where it has a value (the cosmological constant)
\begin{equation}\lb{vm12}
\tilde{V}(\lambda)=\lambda(1+\lambda^5)^{-\frac{1}{5}}~~.
\end{equation}

A profile of the four-dimensional inflaton scalar potential, originating from $D=12$, is given in Fig.~2.

\begin{figure}[htbp] 
\begin{center}
\includegraphics[clip, width=8cm]{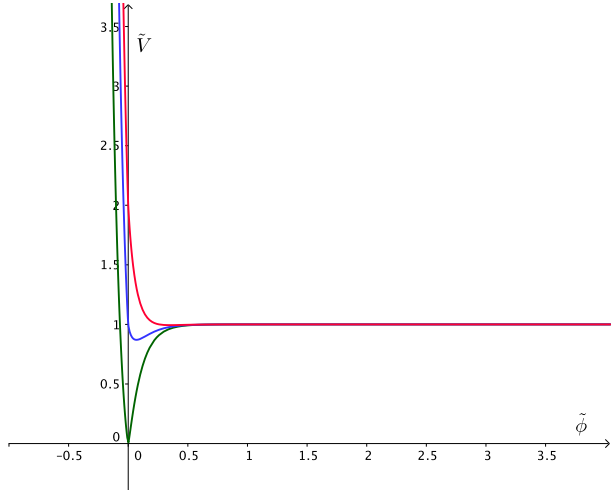}
\caption{The profile of the scalar potential potential obtained from $D=12$ dimensions. The green line 
describes the case of $\lambda=0$, the blue line is of $\lambda=1$, and the red line is of $\lambda=2$, respectively.}
\label{fig:one}
\end{center}
\end{figure}

According to Eq.~(\ref{alphaf}), the parameter $\gamma$ in $D=12$ has the value  $\gamma_{12}\approx\frac{5^4}{2^9\cdot 3^7\cdot 11}10^{492}$. Since its
(mass) dimension is $2-D=-10$, the relevant scale in $D=12$ is given by  
\begin{equation}\lb{scale12}
\gamma_{12}^{-1/10}\approx 1.696\cdot 10^{-49}M_{\rm Pl}~~.
\end{equation} 

\end{itemize}

\section{Conclusion}

In this paper we derived the inflaton scalar potential from higher $(D>4)$ dimensions, in the context of the
$D$-dimensional $(R+\gamma R^n-2\Lambda)$ gravity, by using the Starobinsky model of chaotic large-field inflation in $D=4$ as a prototype. We assumed that a compactification of the extra dimensions took place {\it before} inflation. We found that this requires a positive cosmological constant and $n=D/2$ for consistency. We calculated the corresponding scalar potential and the values of its parameters for any $D$, and specified our results to the two special cases, $D=8$ and $D=12$. Our models predict the viable values
of $n_s$ and $r$ for $N_e$ around $55\div 5$.~\footnote{The four-dimensional inflationary models based on a higher-dimensional $(R+\gamma R^n-2\Lambda)$ gravity were also considered in  Ref.~\cite{zhuk5}, though only for $D<8$ where they were found to be not viable because of low values of $n_s$ and $N_e$. Our results also differ from Ref.~\cite{moto} where only the values of $n$ close to $2$ in four dimensions were studied.}
 
Our scalar potentials in their slow-roll part fall into the class of the inflationary plateau-type potentials describing chaotic large-field inflation and having the form (\ref{pclass}) with
\begin{equation} \lb{slope}
\alpha= \sqrt{ \frac{D-2}{D-1} } ~~,
\end{equation}
because of Eq.~(\ref{4dspot}).  In particular, we have $\alpha_4=\alpha_s=\sqrt{2/3}$,  $\alpha_8=\sqrt{6/7}$ and $\alpha_{12}=\sqrt{10/11}$.

According to Eq.~(\ref{rbound}), the value of the $\alpha$-parameter dictates the observable value of the tensor-to-scalar ratio $r$ as
\begin{equation} \lb{rbound2}
r =\frac{8(D-1)}{(D-2)N_e^2}~~.
\end{equation}
In particular, we find $r_4=\frac{12}{N_e^2}$, $r_8=\frac{28}{3N_e^2}$, and $r_{12}=\frac{44}{5N_e^2}$. All those values are in agreement
with current observations, and give the sharp predictions for future measurements of $r$.

\section*{Acknowledgements}

S.V.K. is supported by a Grant-in-Aid of the Japanese Society for Promotion of Science (JSPS) under No.~26400252, a TMU President Grant of Tokyo Metropolitan University in Japan, the World Premier International Research Center Initiative (WPI Initiative), MEXT, Japan, and the Competitiveness Enhancement Program of Tomsk Polytechnic University in Russia. The authors are grateful to  J.~D.~Barrow, 
H. Motohashi and A. Zhuk for discussions and correspondence.

\end{document}